# Cell vibron polariton in the myelin sheath of nerve

Bo Song[1]* and Yousheng Shu[2]

[1]*School of Optical-Electrical Computer Engineering, University of Shanghai for Science and Technology, Shanghai 200093, P. R. China.*

[2]*State Key Laboratory of Cognitive Neuroscience and Learning & IDG/McGovern Institute for Brain Research, Beijing Normal University, 19 Xinjiekou Wai Street, Beijing 100875, P. R. China.*

*bsong@usst.edu.cn

## Abstract

**Polaritons are arousing tremendous interests in physics and material sciences for their unique and amazing properties, especially including the condensation, lasing without inversion and even room-temperature superfluidity. Herein, we propose a cell vibron polariton (cell-VP): a collectively coherent mode of a photon and all phospholipid molecules in a myelin sheath which is a nervous cell majorly consisting of the phospholipid molecules. Cell-VP can be resonantly self-confined in the myelin sheath under physiological conditions. The observations benefit from the specifically compact, ordered and polar thin-film structure of the sheath, and the relatively strong coupling of the mid-infrared photon with the vibrons of phospholipid tails in the myelin. The underlying physics is revealed to be the collectively coherent superposition of the photon and vibrons, the polariton induced significant enhancement of myelin permittivity, and the resonance of the polariton with the sheath cell. The captured cell-VPs in myelin sheaths may provide a promising way for super-efficient consumption of extra-weak bioenergy and even directly serve for quantum information in the nervous system. These findings further the understanding of neuroscience on the cellular level from the view of quantum mechanics.**

A polariton, as a quasiparticle of light coherently coupled with polar excitation first proposed 60 years ago[1,2], is arousing tremendous interest in physics and material sciences for its unique and amazing properties as well as the potential applications[3-9]. From the beginning of this century, exciton polaritons in microcavities have been deeply explored, leading to the discoveries of the polariton-based condensation[3-4], room-temperature superfluidity[5], and lasing without inversion[6]. Besides these, exciton polaritons were also observed in micron-/submicron-scale organic crystals without any additional microcavity, causing the promising applications in highly integrated photonic devices and chips[7]. Recently, a vibro polariton was reported that was resulted from the coupling of light with disordered polyvinyl acetate (PVAc) molecules in a metal microcavity, and boosted the Raman scattering cross-section by two to three orders of magnitude[8]. It is worth noting that a dimension of micron (or submicron) scale is applied in the cavities or crystals of the studies above, suggesting this scale plays a critical role in these unexpected observations.

The myelin sheath, as a cell of thin-tube structure wrapping the axon fiber of a neuron and as a key evolutionary acquisition of vertebrates[10-14], has features very similar to the previous microcavities and crystals which have induced the unexpected polaritonic properties and applications. The sheath dimensions were the micron scale identical to the previous cases[15-16]. The myelin sheath is usually considered to serve as an electrical insulator for the axon to provide rapid and high-energy-efficient conduction of electrical impulses by forming a highly organized and compactly multilayered thin film[10-12,17-22]. Recently, neuroscientists have been realizing that the myelin sheath may act as far more than an insulator during signal conduction[23]. An infrared (IR) emission was reported to take place in mitochondria located in the axon of the myelinated neuron[24-27], potentially providing IR photons for nerves. Subsequently, based on classical electromagnetic fields, a hypothesis of micron-scalar waveguide was proposed stating that IR light can travel in the myelin sheaths of nerves, serving as signals between neurons in addition to the well-established electrochemical signals[28-29]. Very recently, Liu and his coworkers proposed that energy of the IR-light signal propagating in myelin sheaths was supplied and amplified when crossing the nodes of Ranvier via periodic relay[30]. It is well known that waveguide performance clearly depends on the refractive difference of the core and cladding materials, while the refractivity is closely related to the excitation frequencies of materials. Hence, the myelin sheath, as majorly composed of phospholipid molecules, is more complicated than a normal waveguide when interacting with IR light because the frequencies of molecular vibration modes are mostly in the IR range[31-32]. Moreover, the rate of bio-photon emission was estimated to be ultra-weak, less than 1 photon/ms/neuron[24,28,33-34]. The quantum effect, thus, cannot be ignored and the coupling of a photon with the myelin sheath should be considered on the molecular to cellular levels from the quantum-electrodynamics view.

Herein, we propose a cell vibron polariton (cell-VP) in the myelin sheath of nerve, namely a collectively coherent mode of a photon and all phospholipid molecules in the myelin sheath which is a cell of nervous system and majorly consists of the phospholipid molecules (Fig. 1). Cell-VP can occur and be resonantly self-confined in the sheath under physiological conditions. The observations benefit from the specifically compact, ordered and polar thin-film structure of the myelin sheath[12,17-18,23], and the relatively strong coupling of the mid-IR photon with the vibrons of phospholipid tails in the myelin. The underlying physics is further revealed as the followings. 1) Upon matching of the frequencies of incident photon and lipid vibrons, their collectively coherent coupling causes a quantum superposition of them in the myelin, leading to the formation of polariton. 2) The polariton then enhances the myelin permittivity significantly. 3) Upon the wavelength of polariton is comparable to the thickness of sheath, their resonance makes the polariton standing waves (SWs) become important, and then most of the wave modes become captive due to total internal reflections. It is worth noting that the specific structure above of myelin sheath makes the sheath display a feature of the micron-scale organic crystals[7] in which polaritons can occur without any additional metal microcavity. Moreover, the cell-VP related vibrations in the myelin exist in the form of vibrons, not phonons. The relatively strong coupling of the mid-IR photon with the vibrons provides a possibility for the collective mode forming, although the direct interaction of the vibrons is weak.

**A quantum model built for infrared-photon coupled myelin sheath**

As shown experimentally, the thickness ($d_m$) of the myelin sheath was 1 μm ~ 3 μm, while the perimeter ($C_m$) and length ($L_m$) were 10 μm ~ 40 μm and $10^1$ ~ $10^2$ μm, respectively[15-16]. Interestingly, the myelin sheath thickness is comparable to the wavelength $\lambda_e$ of light emitted by the C-H stretching vibration (Fig. 1 right) in phospholipid tails of myelin[31-32]. Moreover, the energy (0.36 eV) of a photon absorbed by the above C-H vibration[31-32] matches the emitted one (~0.33 eV) per citric acid cycle occurring in mitochondrion locating inside the axon of the myelinated nerve [25-27], with a difference only in the physiologically thermal fluctuation (~0.03 eV). In addition, the interactions between these C-H dipoles are so weak (~$k_B T$ at the physiological temperature $T$ = 37 °C) that their excited vibrations display a single molecular feature in myelin rather than a collective mode (see Sections S1-S2 of the Supplementary Information). This fact indicates that the C-H vibration excitations exist the form of vibrons but not phonons. Based on the characteristics of the myelin sheath, together with the ultra-weak emission of biophotons[24,28,33-34], we built a full quantum mechanics (QM) model involving the photon-vibron coupling for exploring its effects in the myelin sheath.

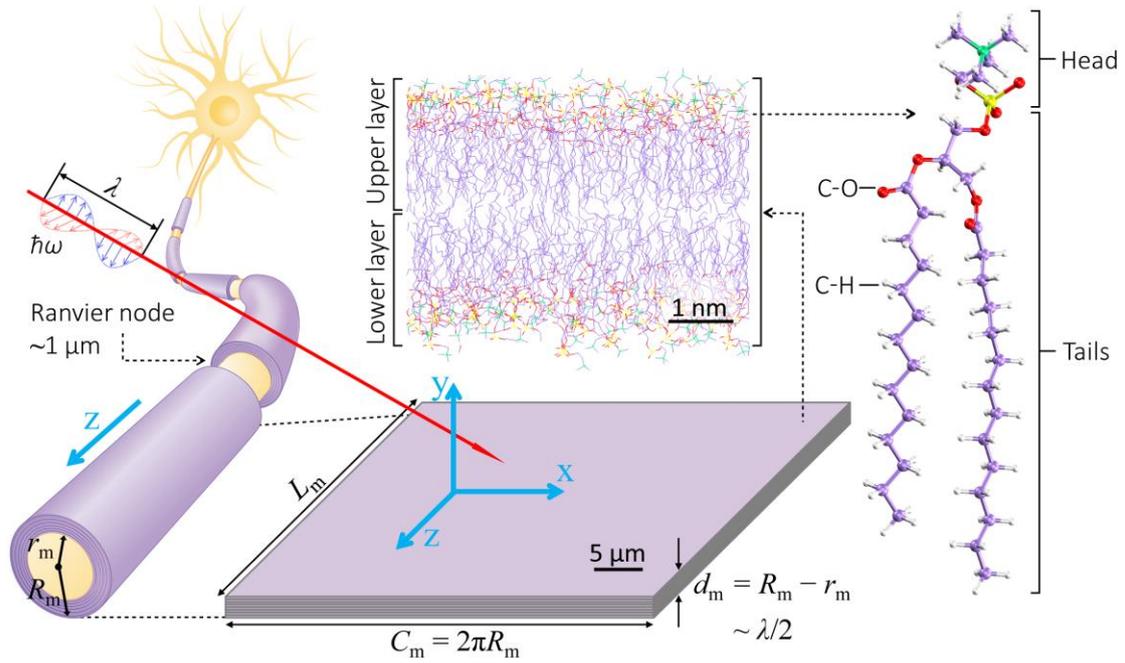

**Figure 1 | Schematic representation of the myelin sheath (a cell, purple tubes) surrounding the axon (another cell, orange cylinder) of the myelinated nerve excited by a mid-infrared phonon $\hbar\omega$ (red arrow) with a specific frequency $\omega$.** The labels $r_m$ and $R_m$ indicate the inside and outside radii, respectively, of the myelin sheath, while $L_m$ denotes the length. The thickness $d_m = R_m - r_m$ is much less than the length ($L_m$) and perimeter ($C_m$) of the sheath. A model of thin film (the lower middle) is thus employed for the sheath to explore the standing-wave effect with a wavelength $\lambda \sim 2d_m$, with a periodic boundary condition applied in the x direction on the y-z surface of the film to better simulate the sheath. Upper middle: molecular structure of a phospholipid bilayer. The myelin sheath consists of $\sim 10^2$ bilayers. Right: a phospholipid molecule with the lipid head, tail, C-O and C-H bonds labeled. The purple, green, red, yellow and white balls indicate the C, N, O, P and H atoms.

Considering that the myelin thickness was much less than its perimeter and length [15-16] and the confinement of specific-frequency light in a cylinder majorly depended on the cylindrical thickness (but not on its diameter)[30], we employed a thin-film model (Fig. 1 middle lower) for myelin sheath to present physics more clearly. To better simulate the sheath, we introduced a periodic boundary condition in the x direction on the y-z surface of the film. According to the experimental data presented before, the model was then built with the dimensions $d_m \sim \lambda_e/2$, $C_m \sim 10\lambda_e$, $L_m \sim 50\lambda_e$. In the rectangular isotropic dielectric film model of the myelin sheath, the Maxwell equation solutions for the electric field were given in the form

$$\mathbf{E}(\mathbf{r},t) = \mathbf{E}_{\mathbf{k},\alpha}(\mathbf{r})\cos(\omega_{\mathbf{k}}t), \quad \varepsilon_{m,r}^0 \omega_{\mathbf{k}}^2 = k^2 c^2, \tag{1}$$

with

$$k = |\mathbf{k}|,$$

where $\varepsilon_{m,r}^0$ denotes the residually relative permittivity of myelin excluding the contribution from the lipid-tail vibrons (whose impact is discussed in detail later). The two polarization vectors $\tilde{\mathbf{u}}_{\mathbf{k},\alpha} = (\tilde{\mathbf{u}}_{\mathbf{k},\alpha}^x, \tilde{\mathbf{u}}_{\mathbf{k},\alpha}^y, \tilde{\mathbf{u}}_{\mathbf{k},\alpha}^z)$ were employed with $\alpha = 1, 2$, with $\tilde{\mathbf{u}}_{\mathbf{k},\alpha} \cdot \tilde{\mathbf{u}}_{\mathbf{k},\alpha'} = \delta_{\alpha,\alpha'}$. The wave functions were thus written as $\mathbf{E}_{\mathbf{k},\alpha}(\mathbf{r}) = (u_{\mathbf{k},\alpha}^x E_{\mathbf{k}}^x(\mathbf{r}), u_{\mathbf{k},\alpha}^y E_{\mathbf{k}}^y(\mathbf{r}), u_{\mathbf{k},\alpha}^z E_{\mathbf{k}}^z(\mathbf{r}))$, where the functions $E_{\mathbf{k}}^x(\mathbf{r})$, $E_{\mathbf{k}}^y(\mathbf{r})$ and $E_{\mathbf{k}}^z(\mathbf{r})$ consist of sine and cosine with respect to $k_x x$, $k_y y$ and $k_z z$ (see Section S3 of the Supplementary Information). There were two independent groups of solutions according to a parallel electric-field boundary condition ($\mathbf{E}_{\mathbf{k},\alpha}(\mathbf{r}_s) \cdot \mathbf{n}_s = 0$) on all surfaces and a parallel magnetic-field boundary condition (where the electric field is normal to the surface: $\mathbf{E}_{\mathbf{k},\alpha}(\mathbf{r}_s) \times \mathbf{n}_s = 0$), respectively, with $\mathbf{n}_s$ denoting the normal of the $s$ surface and $\mathbf{r}_s$ denotes a location on the $s$ surface, with $s = $ x, y, z (see Section S3 of the Supplementary Information). The wave vector $\mathbf{k} = (k_x, k_y, k_z)$ was given by

$$k_x = 2n_1 \pi / C_m, \quad k_y = n_2 \pi / d_m, \quad k_z = n_3 \pi / L_m, \quad n_i = 0, 1, 2, 3, ...$$
$$\tilde{\mathbf{k}} = \mathbf{k} / |\mathbf{k}| = \tilde{\mathbf{u}}_{\mathbf{k},1} \times \tilde{\mathbf{u}}_{\mathbf{k},2}. \tag{2}$$

Hence, a resonant mode of the electric component of the electromagnetic field in the myelin sheath can be a linear combination of the solutions above. The SWs of the solutions were a linear superposition of plane waves with mirror reflections. A wave was thus trapped inside upon the condition of projected wave vector $\mathbf{k}_\parallel$ to the surface

$$\mathbf{k}_\parallel^2 c^2 > \varepsilon_s \omega_{\mathbf{k}}^2, \tag{3}$$

held on all the surfaces, where $\varepsilon_s$ indicates the permittivity of the surrounding media involving axon and extracellular fluid[35]. Note that complex wave vectors could be employed for evanescent waves outside the myelin sheath, and the same case could also be applied to a light not traveling at a resonant frequency inside. In principle, the Maxwell equations can always be solved in a large surrounding space excluding the sheath microcavity, and the couplings of the microcavity with the surroundings can be determined by the boundary continuity of the electromagnetic fields.

To obtain a full QM model of photons interacting with the vibrons, we quantized the macroscopic Maxwell field under the previous eigenmodes of the sheath microcavity[36-37]. For a given $\mathbf{k}$ and boundary conditions, the photon Hamiltonian and electric field operator inside the myelin microcavity were written as

$$H_{ph,k} = \sum_{\alpha} \hbar\omega_k (b^{\dagger}_{k,\alpha} b_{k,\alpha} + \frac{1}{2}),  \quad (4)$$

$$\Pi_k(\vec{r}) = \sum_{\alpha} g_k (b_{k,\alpha} + b^{\dagger}_{k,\alpha}) E_{k,\alpha}(r),  \quad (5a)$$

$$g_k = \sqrt{\hbar\omega_k / (2\varepsilon_0 \varepsilon^0_{m,r})},  \quad (5b)$$

where $b^{\dagger}$ and $b$ denote the creation and annihilation operators of a photon, while $\Pi_k$ indicates the field operator of $E_{k,\alpha}(r)$. The phase of $b_{k,\alpha}$ was chosen for symmetric $\Pi_k$.

The myelin sheath can be illuminated by IR lights emitted from mitochondria[24-27]. The rate of bio-photon emission has been estimated to be extra-weak, less than 1 photon per microsecond per neuron[24,28,33-34]. We thus considered a single vibron limit and approximated an excited C-H dipole in phospholipid tails of myelin by a harmonic oscillator $B^{\dagger}$ ($B$) restricted to its excited level $E_1 = \hbar\omega_e$. Let the location of the $j$th dipole be $R_j$, we had the vibron-photon interaction Hamiltonian as follows

$$H_{int,k} = \sum_{j,\alpha} t_{j,k,\alpha} (B^{\dagger}_j b_{k,\alpha} + b^{\dagger}_{k,\alpha} B_j),  \quad (6)$$

with

$$[B_i, B^{\dagger}_j] = \delta_{i,j}, \quad i,j = 1,2,...,N,$$
$$t_{j,k,\alpha} = g_k \mu_j \cdot E_{k,\alpha}(R_j),$$

where $\mu_j$ indicates the $j$th dipole vector in the phospholipid tails of the myelin sheath. We employed the rotating wave approximation[38-39] when focusing on the resonance processes of the photon and myelin sheath.

The coupling strength between the C-H dipoles of the phospholipid tails in the myelin was ~0.04 eV (see Section S2 of the Supplementary Information), very close to the thermal fluctuation ($k_B T$ ~ 0.03 eV) at the physiological temperature $T = 37$ °C. The interaction between the vibrons was thus not considered in our studies.

Hydrophobic effects of lipid tails can drive the biological self-assembly of phospholipid molecules to form a highly organized lipid bilayer membrane with an ordered structure of the phospholipid molecules[40,41-46]. The myelin sheath consists of hundreds of the above lipid bilayers with a compact and quasi-periodic structure (Fig. 1 middle)[15,47-48]. Moreover, there exists anisotropy of the C-H dipoles as well as their excitations of lipid tails in myelin sheath (see Section S4 of the Supplementary Information), making the myelin display a profile of polar film[23,45-46,49]. All of these facts suggest that the myelin sheath presents a feature of organic crystals, although the coupling of dipoles in the lipid tails is too weak (~$k_B T$, see Section S2 of Supplementary Information) to induce a collective excitation mode like the phonon at physiological temperature. Following the approaches in the literatures[39,50], the vibron can be rewritten in an SW superposition of the identical states from all the phospholipid tails in the myelin via a unitary transformation,

$$B_{z,\mathbf{k}}^{\dagger} = \sqrt{\frac{V}{N}} \sum_j U_{z,\mathbf{k}}(\mathbf{R}_j) B_j^{\dagger},$$

$$|\varphi_{z,\mathbf{k}}\rangle = B_{z,\mathbf{k}}^{\dagger}|0\rangle, \quad [B_{z,\mathbf{k}}, B_{z,\mathbf{k}'}^{\dagger}] = \delta_{\mathbf{k},\mathbf{k}'}.$$

where $N$ indicates the number of the vibrons in the myelin sheath, and $V$ denotes the volume of the sheath. Likewise, we also defined the SW states $|\varphi_{x,\mathbf{k}}\rangle$ and $|\varphi_{y,\mathbf{k}}\rangle$ for the x and y components, respectively. The three SW states are mutually orthogonal for a large $N$.

In terms of the coherent SW states[39,50], the total Hamiltonian of photon-vibron coupling system was given as follows

$$H = \sum_{\mathbf{k}} H_{0,\mathbf{k}} + H_{int,\mathbf{k}},$$

where the zeroth order term is

$$H_{0,\mathbf{k}} = \sum_{\gamma} \hbar\omega_e B_{\gamma,\mathbf{k}}^{\dagger} B_{\gamma,\mathbf{k}} + \sum_{\alpha} \hbar\omega_{\mathbf{k}} b_{\mathbf{k},\alpha}^{\dagger} b_{\mathbf{k},\alpha}, \quad \gamma = x, y, z,$$

and the photon-vibron interaction of equation (5) is rewritten as

$$H_{int,\mathbf{k}} = \sum_{\alpha,\gamma} \left[ t_{\gamma,\mathbf{k},\alpha} (B_{\gamma,\mathbf{k}}^{\dagger} b_{\mathbf{k},\alpha} + b_{\mathbf{k},\alpha}^{\dagger} B_{\gamma,\mathbf{k}}) \right],$$

$$t_{\gamma,\mathbf{k},\alpha} = g_{\mathbf{k}} \boldsymbol{\mu}_{\gamma} \cdot \tilde{\mathbf{u}}_{\mathbf{k},\alpha} \sqrt{N/V} \equiv t_{\mathbf{k}} \tilde{\boldsymbol{\mu}}_{\gamma} \cdot \tilde{\mathbf{u}}_{\mathbf{k},\alpha},$$

$$t_{\mathbf{k}} = g_{\mathbf{k}} |\boldsymbol{\mu}| \sqrt{N/V}, \quad \tilde{\boldsymbol{\mu}}_{\gamma} = \boldsymbol{\mu}_{\gamma} / |\boldsymbol{\mu}|,$$

with $\boldsymbol{\mu}_{\gamma}$ indicating the projection of the dipole vector $\boldsymbol{\mu}$ in the $\gamma$ direction. The dispersion of the resulted quantum superposition state of cell-VP was then obtained as follows

$$E_{\mathbf{k}}^{(\pm)} = \frac{\hbar\omega_e + \hbar\omega_{\mathbf{k}}}{2} \pm \sqrt{\left(\frac{\hbar\omega_e - \hbar\omega_{\mathbf{k}}}{2}\right)^2 + \frac{t_{\mathbf{k}}^2}{3}}. \tag{7}$$

**Results and discussions**

We are interested in whether a cell-VP captured in the myelin sheath exists under physiological conditions. The emitted energy per citric acid cycle occurring in axonal mitochondrion is ~32 kJ/mol (0.33 eV)[25-27], equal to the energy of a mid-IR photon with a frequency of 80 THz. The frequency 87 THz (0.36 eV) of the C-H stretching vibration in lipid tails (Fig. 1 right)[31-32] is very close to the one above of the mid-IR photon. Hence, we employed an energy $\hbar\omega_e = 0.36$ eV of the photon emitted by the myelin in our following calculations, corresponding to the frequency (87 THz) of the C-H vibration in $CH_2$ groups of the lipid tails. The previously experimental reports of visible light suggested that relative permittivity of neuron was ~2.07 for the myelin sheath, ~1.90 for the axon, and ~1.80 for the interstitial fluid outside[35]. Excluding the cell-VP contribution, we thus set $\varepsilon_{m,r}^{0} = \varepsilon_{s,r} = 1.85$, namely, an extreme case that the residually relative permittivity of myelin was identical to the surrounding media. Therefore, under this permittivity of myelin, the wavelength of the emitted 87-THz photon is determined as $\lambda_e = 2.5$ μm for the case without the quantum resonance and coherence of photon and vibrons. Further considering the experimental data shown previously[15-16], we applied the myelin sheath parameters $d_m = 1.25$ μm ($= \lambda_e/2$), $C_m = 25$ μm ($= 10\lambda_e$) and $L_m = 125$ μm ($= 50\lambda_e$) for simplicity. Based on the experimental and *ab initio* calculation data, considering the cerebrospinal-fluid pressure on the lipid density, we obtained a coupling strength $t_k \approx 0.18$ eV ($= 0.5\hbar\omega_e$) of the 87-THz photon with the C-H dipole in lipid tails of the myelin sheath (see Section S5 of the Supplementary Information).

**The cell vibron polariton forming in the myelin sheath at physiological temperature**. As shown in Fig. 2a, there were two branches in the cell-VP dispersion, which was attributed to the quantum coherent superposition of photon and vibrons. For a large wave vector $k$, the behavior of the upper branch was similar to the photon, and the lower branch was similar to the vibron. To explore the thermal stability of cell-VPs, we employed an energy drop $\Delta_k$ of the lower branch (cell-VP ground state) with respect to the uncoupled vibron and photon as follows,

$$\Delta_k = \begin{cases} \hbar\omega_e - E_k^{(-)}, & \text{when } \hbar\omega_e \leq \hbar\omega_k \\ \hbar\omega_k - E_k^{(-)}, & \text{when } \hbar\omega_e > \hbar\omega_k \end{cases}$$
$$= -\frac{|\hbar\omega_e - \hbar\omega_k|}{2} + \sqrt{\left(\frac{\hbar\omega_e - \hbar\omega_k}{2}\right)^2 + \frac{t_k^2}{3}}. \quad (8)$$

The results under the above biological conditions are presented in Fig. 2c. When $k = k_e$ ($= 2\pi/\lambda_e$) (the optimal coherence point), $\Delta_k = 0.29\hbar\omega_e$ ($= 0.10$ eV) was clearly larger than the thermal fluctuation $k_B T \sim 0.03$ eV ($T = 37$ °C), denoting a more stable superposition state cell-VP than the uncoupled photon or vibron at physiological temperature. Moreover, the cell-VP energy $E_k = 0.71\hbar\omega_e$ ($= 0.26$ eV) at the point $k = k_e$ of optimal coherence was much larger than the thermal fluctuation. Therefore, cell-VP can reach a relatively stable state with an ignorable lifetime. As $k$ increased, the energy drop $\Delta_k$ decreased and was less than $k_B T$ when $k > 1.91 k_e$, which indicates that the cell-VP state is unstable for a large wave vector at physiological temperature. We have also studied the effect of photon-vibron coupling strength $t_k$ on the cell-VP stability. As shown in Fig. 2c right, only when $t_k < 0.17\hbar\omega_e$ ($= 0.06$ eV), the maximum $\Delta_{max}$ of the energy drop $\Delta_k$ was less than $k_B T$, indicating that cell-VPs cannot form at physiological temperature only when the photon-vibron coupling is very weak. Therefore, a collectively coherent mode cell-VP can be caused by the relatively strong coupling (~0.18 eV) of photon with dipoles of the phospholipid tails in myelin, although the direct interaction between the related C-H dipoles of lipid tails is too weak (~$k_BT$, see Section S2 of the Supplementary Information) to induce a phonon-like excitation mode at physiological temperature.

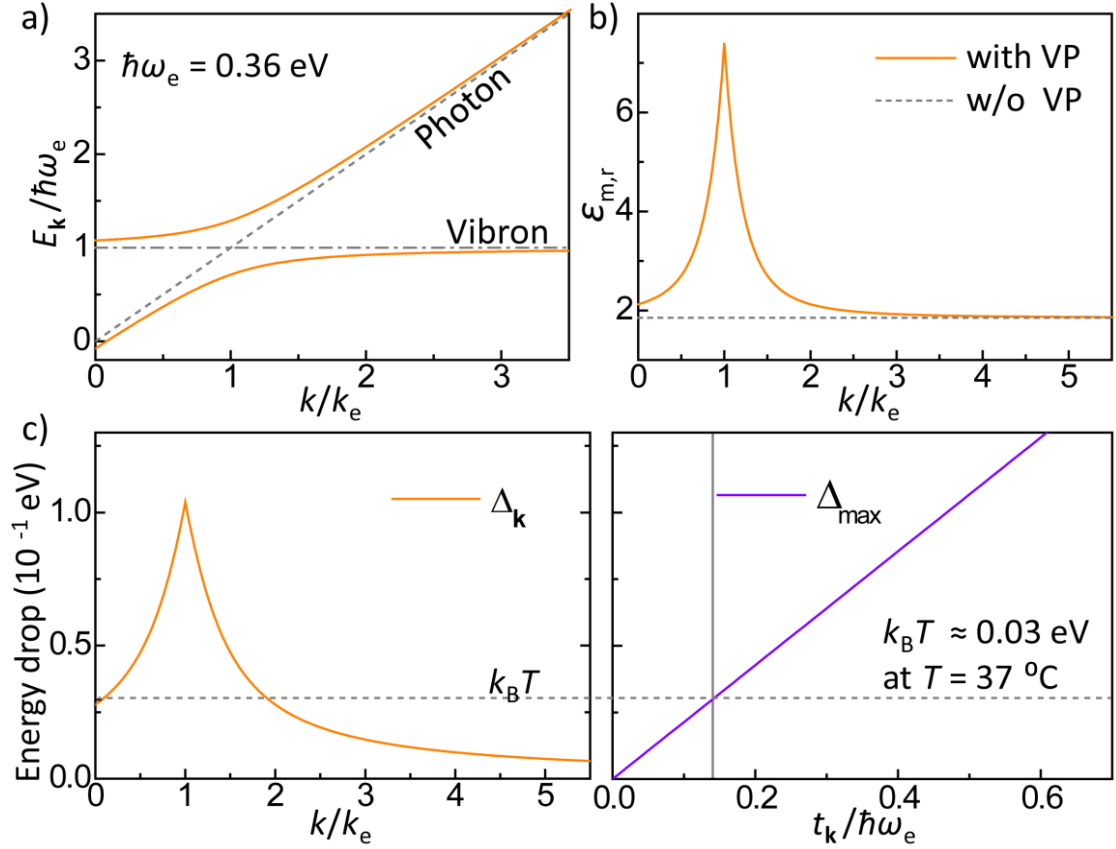

**Figure 2 | Formation of cell vibron polariton (cell-VP) in the myelin sheath under physiological conditions as well as its impact on myelin permittivity. a)** Dispersion of cell-VP. The orange curves indicate two branches of cell-VP dispersion. The dashed and dot-dashed lines denote the dispersions of uncoupled photon and vibron, respectively. For C-H stretching vibrons of the phospholipid tails in the myelin sheath of a neuron, $t_k = 0.5\hbar\omega_e$ and $\hbar\omega_e = 0.36$ eV. **b)** Comparison of the relative permittivity $\varepsilon_{m,r}$ of myelin with (orange solid) and without (silver dotted) the effect of cell-VP. **c)** Energy drop of the cell-VP lower branch with respect to uncoupled photon and vibron. The orange and violet curves indicate the energy drop $\Delta_k$ to $k$ (left) and its maximum value $\Delta_{max}$ to $t_k$ (right), respectively. The dashed line denotes the value of thermal fluctuation $k_BT \sim 0.03$ eV at the physiological temperature $T = 37$ °C. Right: The vertical line represents the vibron-photon coupling strength $t_k$ at $\Delta_{max} = k_BT$.

**The formed cell vibron polariton being captured within the myelin sheath**. As a qualitative analysis, we first explored the impact of cell-VP formation on the permittivity of myelin. The relative permittivity $\varepsilon_{m,r}$ of myelin sheath including the cell-VP contribution was calculated based on the dispersion of equation (7) (see details in Section S6 of Supplementary Information), and the results are shown in Fig. 2b. A peak of 7.40 was observed at the point of optimal coherence $k = k_e$, much higher than the value 1.85 (i.e., $\varepsilon_{m,r}^0$) without the cell-VP effect. The permittivity then decreased as the wave vector $k$ increased or decreased from $k_e$. For a large $k$, $\varepsilon_{m,r}$ slowly dropped back to $\varepsilon_{m,r}^0$. Therefore, in the coherent coupling area, the formed cell-VP can significantly enhance the myelin permittivity, providing a critical base for the sheath to trap the light even if myelin permittivity excluding cell-VP effects is initially set as identical to the surroundings. Considering the relation of permittivity and refractive index together with the definition of the index, we note that the enhancement of permittivity is physically related to the slowing down of photon due to resonant absorption and emission of the photon by phospholipid molecules in the compact myelin. The slowing-down rate depends on the phospholipid-photon coupling strength. The total internal reflection criterion of equation (3) was then employed for our quantitative studies, from which a dimensionless parameter $\delta_{\mathbf{k},\|}$ was introduced as

$$\delta_{\mathbf{k},\|} = \mathbf{k}_\|^2 / k_e^2 - \varepsilon_{s,r} E_\mathbf{k}^2 / (\hbar k_e c)^2 > 0, \tag{9}$$

with $\mathbf{k}_\|^2 = (k_x^2 + k_y^2), (k_y^2 + k_z^2), (k_z^2 + k_x^2)$, $k_e = \sqrt{\varepsilon_{m,r}^0} \cdot \omega_e / c = 2\pi / \lambda_e$, and $E_\mathbf{k} = E_\mathbf{k}^{(-)}$ in equation (7). Biologically considering the tube structure of the myelin sheath (Fig. 1 left), we ignored the criterion on the y-z surface, and just studied the cases simply with $k_x = 0$, which gave $\mathbf{k}_\|^2 = k_y^2$ for the x-y surface and $k_z^2$ for the z-x surface. The results under the physiological conditions are shown in Fig. 3a-b. The criterion was reached simultaneously on the x-y and z-x surfaces when $k_z > 0.82 k_e$ for $k_y = k_e$, and $k_z > 0.93 k_e$ for $k_y = 2 k_e$. These facts indicate that cell-VP is able to undergo total internal reflections on all surfaces of the myelin sheath at the same time. Additionally, the condition of cell-VP thermal stability ($\Delta_k > k_B T$) at physiological temperature should also be considered, which gave the restrictions (Fig. 3c): $k_z < 1.63 k_e$ upon $k_y = k_e$, and no $k_z$ upon $k_y = 2 k_e$. Hence, there exists a window ($k_x = 0$, $k_y = k_e$, $0.93 k_e < k_z < 1.63 k_e$) for cell-VPs captured inside the microcavity of the myelin sheath under physiological conditions. Note that the SWs in the sheath employed as the basis set of our calculations further provided a resonant confinement of the cell-VP wave in the microcavity of the sheath with quick attenuation outside. Additionally, the periodic boundary condition in the x direction makes us able to ignore the criterion of total internal reflection on the y-z surface, leading to that $k_x$ can be used to further regulate the confinement of cell-VP.

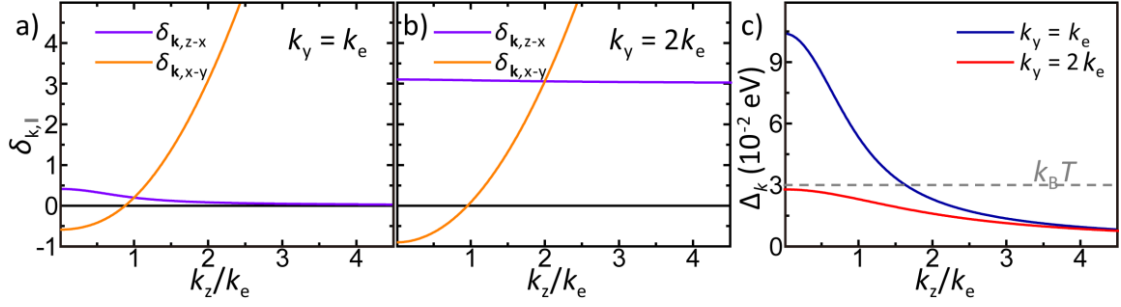

**Figure 3 | Capture of cell vibron polariton within the myelin sheath under physiological conditions. a-b)** Total internal reflection criterion of cell-VP $\delta_{\mathbf{k},\|} = \mathbf{k}_\|^2 / k_e^2 - \varepsilon_{s,r} E_\mathbf{k}^2 / (\hbar k_e c)^2 > 0$. The violet and orange curves denote the cases of $\mathbf{k}_\|^2$ = ($k_z^2 + k_x^2$) and ($k_x^2 + k_y^2$), respectively. The wave vector $k_y = k_e$ (a) and $2k_e$ (b). **c)** Energy drop of cell-VP with respect to the uncoupled photon and vibron. The silver dashed line represents the thermal fluctuation $k_B T$ (~ 0.03 eV) at the physiological temperature $T = 37$ °C. $k_x = 0$ is applied to present the physics clearly.

It is worth noting that identical observations can also be expected for C=O stretching (Fig. 1 right) of phospholipid molecules (vibration frequency $\nu_{C=O} = 52$ THz [30-31]) in myelin, when the incident-photon frequency $\nu \approx \nu_{C=O}$ and the sheath thickness $d_m \sim 2.9$ μm ($0.5\lambda_{C=O}$). Moreover, the methylene bridge and carbonyl group are enriched compactly and orderly in protein fiber bundles of cytoskeleton[27,45], which also potentially causes cell-VP upon the dimension of cell involving neuron comparable to the incident mid-IR wavelength[45,46].

Based on the previous calculations and analyses, we obtained a picture of mid-IR light resonantly captured by the myelin sheath of a nerve, which potentially occurs *in vivo*. For an incident photon with a wavelength $\lambda$ comparable to $d_m$ and a frequency $\nu \approx \nu_{myelin}$ (in the mid-IR range), the following conditions

$$\Delta_\mathbf{k} > k_B T, \ E_\mathbf{k} > k_B T \text{ and } \delta_{\mathbf{k},\|} > 0 \tag{10}$$

can be simultaneously satisfied in the myelin sheath under physiological conditions. This fact indicates that the energy of the special mid-IR photon can be stored in cell-VP, which can be resonantly trapped in the myelin sheath *in vivo*. Note that leaking of the captive modes through the edges of myelin sheath is still possible but should be very weak (the magnitude ~ $N^{-2/3}$ << 1 for the edges-to-bulk ratio, $N$ being the number of C-H dipoles in the sheath). Remarkably, with help of quantum resonance, the energy of cell-VP can be accepted by a channel protein in the Ranvier node (Fig. 4, Fig. 1 left) if they have the identical eigen-frequency. This accepted energy may be consumed by the protein to promote the conduction of electric impulses in the neuron. Moreover, the cell-VP can also hop between myelin sheaths through quantum tunneling, which may directly serves for quantum information of nerves. It is worth noting that due to the polaritons majorly leaking through myelin edges, the cylindrical structures of myelin sheaths axially connected along an axon (Fig. 1 left) provide a promising way for the polaritonic quantum information transmission between the sheaths. These potential functions of cell-VPs in the nervous system are beyond this work and should be explored in further studies.

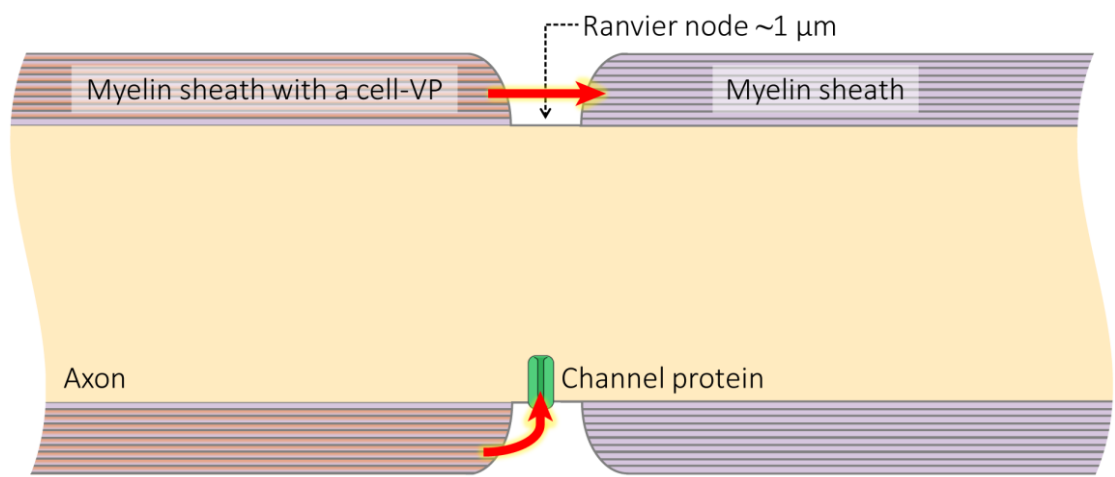

**Figure 4 | Schematic representation of the longitudinal section of myelinated axon with a cell-VP excitation in the left sheath and its potential functions in the nervous system.** The Ranvier node is a gap (~1 μm) between two myelin sheaths, in which the channel protein (green) locates. The upper red arrow indicates the potential quantum tunneling of cell-VP between two sheaths, and the lower one denotes the potential quantum-resonance energy transfer from cell-VP in the sheath to the vibration of channel protein in the Ranvier node.

**Conclusion**

We proposed a cell vibron polariton forming and captured in the myelin sheath of nerve under physiological conditions, when the incident photon frequency matched the lipid vibration frequency, and its wavelength was comparable to the sheath thickness. The observations benefited from the specifically compact, ordered and polar thin-film structure of the myelin sheath, and the relatively strong coupling of the mid-IR photon with the phospholipid-tail vibrons in the myelin. The underlying physics was clearly revealed to be the collectively coherent superposition of the photon and all the vibrons, the formed polariton induced significant enhancement of myelin permittivity, and the resonance of the polariton with the sheath cell. We note that with help of further quantum resonances, the captured cell-VPs can provide an expectable way for super-efficient consumption of extra-weak bioenergy, and even may directly serve for quantum information to the nervous system. These findings enhance the understanding of biology from a quantum and cellular view.

**Acknowledgements**

B.S. thanks Dr. Bin Cheng for the fruitful discussions. This work was supported by the National Natural Science Foundation of China Projects (31630029, 31661143037) and the National Supercomputer Center in Tianjin, and dedicated to the memory of B.S.'s mother.

**Author contributions**

B.S. contributed to the idea. The theoretical calculations and analyses were designed and carried out by B.S.. Y.S. and B.S. performed neurobiological analyses. B.S. and Y.S. co-wrote the paper. All authors discussed the results and commented on the manuscript.

**Competing interests**

The authors declare no competing interests.